\input preprint.sty

\def\dpr{{\ooalign{\hfil\raise.07ex\hbox{$\times$}\hfil\crcr\mathhexbox20D}}}
\def\sdp{{\ooalign{\hfil\raise.07ex\hbox{s}\hfil\crcr\mathhexbox20D}}}
\def\append#1
   {\vskip0pt plus.1\vsize\penalty-250
    \vskip0pt plus-.1\vsize\vskip24pt plus12pt minus6pt
    \subno=0
    \global\advance\appno by 1
    \noindent {\bf Appendix. #1\par}
    \bigskip
    \noindent}

\title{Critical temperature of a fully anisotropic three-dimensional
       Ising model}

\author{M A Yurishchev}

\address{Radiophysical Research Institute,
Nizhny Novgorod 603950, Russia}

\shorttitle{Critical temperature of Ising model}



\date

\beginabstract
The critical temperature of a three-dimensional Ising model on a
simple cubic lattice with different coupling strengths along all
three spatial directions is calculated via the transfer matrix
method and a finite size scaling for $L\times L\times\infty$
clusters ($L=2$ and 3).
The results obtained are compared with available calculations.
An exact analytical solution is found for the
$2\times2\times\infty$ Ising chain with fully anisotropic
interactions (arbitrary $J_{\rm x}$, $J_{\rm y}$ and $J_{\rm z}$).
\endabstract

\section{Introduction}
Great attention is given to the critical temperature
calculation of the three-dimensional Ising model,
for the full time extent of existence.
The most considerable advances have been attained for the
fully isotropic cubic lattice ($J_{\rm x}=J_{\rm y}=J_{\rm z}$).
The  calculations are steadily improving with time, accuracy
for the critical point value is down to $10^{-3}\%$ error:
$K_{c}=0.221\,6595\pm0.000\,0026$ (Ferrenberg
and Landau 1991), $K_{c}=0.221\,6544\pm0.000\,0010$ (Livet 1991).

The efficiency of progress is less for the partly anisotropic model
when $J_{\rm x}=J_{\rm y}\neq J_{\rm z}$.
Two cases exist here: $J_{\rm x}=J_{\rm y}\geq J_{\rm z}$ (a
quasi-two-dimensional model) and, inversely,
$J_{\rm x}=J_{\rm y}\leq J_{\rm z}$ (a quasi-one-dimensional
model).
Using the high-temperature series, the critical temperature
estimates have been obtained with $2\times10^{-2}\%$ error
for the fully isotropic interactions ($J_{\rm x}=J_{\rm y}=J_{\rm z}$)
and about $10^{-3}\%$ in the two-dimensional limit ($J_{\rm z}=0$)
of quasi-two-dimensional model (see Navarro and de Jongh 1978
and references cited there).
By intermediate range of interlayer couplings, the error of a
phase transition temperature determination lies between these
two extreme values.
Conversely,
in the quasi-one-dimensional case, the estimates based on the
same high-temperature series are rapidly deteriorated
due to the limited number of terms available in the series.
As a result, one can find the critical temperatures only
up $J_{\rm x(y)}/J_{\rm z}=10^{-2}$ (with exactness
in the one-two significant figure range).
In the quasi-one-dimensional case, the phase transition temperature
has been calculated also by phenomenological renormalisation of
clusters (Yurishchev and Sterlin 1991).
Inasmuch, the cluster geometry reflects the physical situation;
this approach (contrary to the high-temperature series
expansions) yields more precise results as the anisotropy
of quasi-one-dimensional system increases.
By $J_{\rm x(y)}/J_{\rm z}=10^{-3}$, the critical temperature
position is determined here with an accuracy of about three
significant figures.

The difficulties are largest when the interactions are different
along all three directions.
Here there are the calculations done
by real-space renormalisation group
method (da Silva \etal 1984)
and the calculations carried out via
various versions of mean field theory and variational principle
(see Faleiro Ferreira 1988, 1989 and references therein).
These results we discuss in detail in the third section
in comparing with our computations.

In this paper, the critical temperature of a three-dimensional
Ising model, with fully anisotropic interactions, is calculated
by the transfer matrix approach in combination with a finite
size scaling, i.e. by the phenomenological renormalisation
group method proposed by Nightingale (1976).
As the clusters, we exploit the infinitely long parallelepipeds
$L\times L\times\infty$ with transverse scales $L=2$ and 3.
We succeeded in obtaining a rigorous solution for lattice
with $L=2$.
For the $3\times 3\times\infty$ lattice, we simplify the partial
eigenvalue problem for the transfer matrix of 512th order.
Using the symmetry, we reduce this problem to a determination
of the largest eigenvalues of the two 18th order matrices.
Final calculations have already been made numerically.

\section{Calculation of the critical temperature}
Conforming with the phenomenological renormalisation group
theory, the critical temperature $T_{\rm c}$ is a fixed point
of an equation (see, for instance, reviews of
Nightingale 1982 and Barber 1983):
$$
\indeqn{L\kappa_L(T_{\rm c})
=L^\prime\kappa_{L^\prime}(T_{\rm c})\eq(1)}
$$
where
$$
\indeqn{\kappa_L=\ln(\lambda_1/\lambda_2)\eq(2)}
$$
is the inverse correlation length in a cluster with characteristic
size $L$.
The quantities $\lambda_1$ and $\lambda_2$ entering into (2) are
respectively largest and next largest eigenvalues of subsystem
transfer matrix.
Thus, the task is reduced to find the dominant
eigenvalues of transfer matrices.

\subsection{Cluster $2\times 2\times\infty$}
Let us write the Hamiltonian for the cluster as
$$
\eqalignno{H&=-\sum_i[J_{\rm x}(\sigma_{1,i}
\sigma_{2,i}+\sigma_{3,i}\sigma_{4,i})
+J_{\rm y}(\sigma_{1,i}\sigma_{4,i}+\sigma_{2,i}\sigma_{3,i})\cr
&\phantom{=}{}+J_{\rm z}(\sigma_{1,i}\sigma_{1,i+1}
+\sigma_{2,i}\sigma_{2,i+1}+\sigma_{3,i}\sigma_{3,i+1}
+\sigma_{4,i}\sigma_{4,i+1})].&(3)\cr}
$$
The spin variables $\sigma=\pm 1$ are located in sites of a lattice
$2\times2\times\infty$ which has a rectangular cross section and
has the symmetry planes going through its axis and the
middles of opposite sides.

The transfer matrix $\bi{U}$ with elements
$$
\eqalignno{\langle\sigma_1,\sigma_2,\sigma_3,\sigma_4|\bi{U}|
\sigma_1^\prime,\sigma_2^\prime,\sigma_3^\prime,\sigma_4^\prime
\rangle&=
\exp[\case{1}{2}K_{\rm x}(\sigma_1\sigma_2+
\sigma_3\sigma_4+\sigma_1^\prime\sigma_2^\prime+
\sigma_3^\prime\sigma_4^\prime)\cr&\phantom{=}{}
+\case{1}{2}K_{\rm y}(\sigma_1\sigma_4+\sigma_2\sigma_3+
\sigma_1^\prime\sigma_4^\prime+\sigma_2^\prime\sigma_3^\prime)\cr
&\phantom{=}{}
+K_{\rm z}(\sigma_1\sigma_1^\prime+\sigma_2\sigma_2^\prime+
\sigma_3\sigma_3^\prime+\sigma_4\sigma_4^\prime)]&(4)\cr}
$$
where $K_{\rm x}=J_{\rm x}/kT$, $K_{\rm y}=J_{\rm y}/kT$ and
$K_{\rm z}=J_{\rm z}/kT$ corresponds to the Hamiltonian (3).

To solve the eigenvalue problem of the transfer
matrix (4), we use first the invariance property of the appropriate
Hamiltonian with respect to the transformations of the group
$\bss{Z}_2\dpr\bss{C}_{2\rm v}$; where $\bss{Z}_2$ is a group of
global reflections in the spin space, $\bss{C}_{2\rm v}$ is the
point group generated by the symmetry planes of a lattice and
$\dpr$ represents the direct product.
Carrying out the usual group-theoretical analysis (see, e.g.,
Yurishchev 1989), we come to a conclusion that the $16\times16$
transfer matrix (4) can be reduced due to
symmetry $\bss{Z}_2\dpr\bss{C}_{2\rm v}$
to a quasi-diagonal form
with one subblock $5\times5$, four subblocks $2\times2$ and
three `subblocks' $1\times1$, i.e. ready-made eigenvalues.

The subblock of size $5\times5$ is connected with the fully
symmetrical irreducible representation of the group.
Due to the Perron theorem, it contains the largest
eigenvalue of $\bi{U}$.
Basis vectors for this irreducible representation are given as
$$
\indeqn{\psi_1=\frac{e_1+e_{16}}{\sqrt{2}}\qquad
\psi_2=\frac{e_2+e_3+e_5+e_8+e_9+e_{12}+e_{14}+e_{15}}{2\sqrt{2}}}
$$
$$
\eqno(5)
$$
$$
\indeqn{\psi_3=\frac{e_4+e_{13}}{\sqrt{2}}\qquad
\psi_4=\frac{e_6+e_{11}}{\sqrt{2}}\qquad
\psi_5=\frac{e_7+e_{10}}{\sqrt{2}}}
$$
where
$$
e_1=|1,1,1,1\rangle\qquad e_2=|1,1,1,-1\rangle\qquad\ldots
\qquad e_{16}=|-1,-1,-1,-1\rangle.\eqno(6)
$$
Using these basis functions and utilizing (4), we find the matrix
elements $\psi_i^+\bi{U}\psi_j$ of subblock $5\times5$.
The secular equation of this subblock has a structure
(and this is the
second key circumstance allowing the solution of the
eigenvalue problem):
$$
\indeqn{\lambda^5-a_1\lambda^4+a_2\lambda^3-\alpha a_2\lambda^2
+\alpha^3a_1\lambda-\alpha^5=0.\eq(7)}
$$
Here
$$
\indeqn{a_1=2[1+4\cosh(2K_{\rm x})\cosh(2K_{\rm y})]
\cosh(4K_{\rm z})+6\eq(8)}
$$
$$
\eqalignno{a_2&=32\cosh(2K_{\rm x})\cosh(2K_{\rm y})
[\cosh(4K_{\rm  z})\cosh^2(2K_{\rm z})-1]\cr
&\phantom{=}{}+8[1+\cosh(4K_{\rm x})+\cosh(4K_{\rm y})]
\sinh^2(4K_{\rm z})&(9)\cr}
$$
and
$$
\indeqn{\alpha=4\sinh^2(2K_{\rm z}).\eq(10)}
$$
According to Sominskii (1967), an algebraic equation like (7) is
a reciprocal one.
This property makes it possible to easily find the roots of our
equation.
As a result, the largest eigenvalue of the
transfer matrix (4) is equal to
$$
\indeqn{\lambda_1=\case{1}{2}r_1+(\case{1}{4}r_1^2
-\alpha^2)^{1/2}\eq(11)}
$$
with
$$
\indeqn{r_1=\case{1}{2}(a_1-\alpha)+[\case{1}{4}(a_1+\alpha)^2
+\alpha^2-a_2]^{1/2}.\eq(12)}
$$

Solving secular equations of second-order subblocks causes
no difficulties.
In this issue, we obtain a complete set of eigenvalues.
Sorting the eigenvalues, we seek out the next largest eigenvalue
of $\bi{U}$:
$$
\eqalignno{\lambda_2&=\{1+\exp[2(K_{\rm x}+K_{\rm y})]\}
\sinh(4K_{\rm z})
+\lbrack\!\lbrack\{1-\exp[2(K_{\rm x}+K_{\rm y})]\}^2
\sinh^2(4K_{\rm z})\cr
&\phantom{=}{}+16\exp[2(K_{\rm x}+K_{\rm y})]
\sinh^2(2K_{\rm z})\rbrack\!\rbrack^{1/2}.&(13)\cr}
$$
Note that it lies in the subblock built on basis functions which
are symmetrical under all purely spatial transformations of the
group and antisymmetrical under those including the spin inversion.

By $J_{\rm x}=J_{\rm y}$, our solution is reduced
to that of Kaufman (1949) for the Ising model on a cylinder,
if the number of chains in the last model
is equal to four.

It is also interesting to note that
the above mention does not succeed in generalizing the model (3).
All attempts in including in the Hamiltonian new interactions
(e.g., external field,
additional pair couplings or multiparticle forces) lead
immutably to the destruction of the obvious symmetry of
$\bss{Z}_2\dpr\bss{C}_{2\rm v}$ or the hidden algebraic one
(i.e., the reciprocal property of a secular equation).

\subsection{Cluster $3\times3\times\infty$}
We shall consider a subsystem $3\times3\times\infty$ with cyclic
boundary conditions in both transverse directions.
This eliminates undesirable surface effects and at the same time
extends the symmetry group down to
$\bss{Z}_2\dpr(\bss{T}\sdp\bss{C}_{2\rm v})$, where $\bss{T}$ is
a group of transverse translations and $\sdp$ implies a semidirect
multiplication.
The given symmetry allows one to reduce the transfer matrix
$\bi{V}$, of the size $512\times512$, to a block diagonal form in
which the first and second largest eigenvalues of original matrix
are located in different subblocks ($\bi{V}^{(1)}$ and
$\bi{V}^{(2)}$, respectively) both having a dimension of
$18\times18$.
The open form of these subblocks is given in the appendix.
The extraction of dominant eigenvalues from $\bi{V}^{(1)}$ and
$\bi{V}^{(2)}$ has been carried out already by the computer.

$ $

We return again to the calculation of critical temperature.
The estimates $kT_{\rm c}/J_{\rm z}$ obtained by a numerical
solution of transcendental equation (1) are collected in table 1.
By this, we also put the cyclic boundary conditions on the cluster
$2\times2\times\infty$, i.e.
simply increase the interaction
constants in transverse directions by two times:
$J_{\rm x}\to2J_{\rm x}$ and
$J_{\rm y}\to2J_{\rm y}$.
In table 1, we have also inserted the critical
temperature values for two limited cases: (i) $J_{\rm y}=0$,
corresponding to the anisotropic two-dimensional Ising model
for which the exact phase transition temperature equation
is known (Onsager 1944)
$$
\indeqn{\sinh\left(\frac{2J_{\rm x}}{kT_{\rm c}}\right)
\sinh\left(\frac{2J_{\rm z}}{kT_{\rm c}}\right)=1\eq(14)}
$$
and (ii) $J_{\rm x}=J_{\rm y}$, corresponding to the partly
anisotropic three-dimensional Ising model for which there exists
sufficiently accurate estimates of $T_{\rm c}$ (Navarro and
de Jongh 1978, Yurishchev and Sterlin 1991).

\section{Discussion}
One of simplest ways in estimating the phase transition temperature
in an Ising model is the mean field approximation (MFA):
$$
\indeqn{\left(kT_{\rm c}\right)_{\rm MFA}=2(J_{\rm x}+J_{\rm y}
+J_{\rm z}).\eq(15)}
$$
However, the accuracy is quite low (see table 2 where, for
convenience of comparison, the critical temperature estimates
found by various approximate methods have been given, as well as
the true values obtaining from a solution of equation (14) and
the precision numerical values).

The state of things is somewhat corrected by an improved mean
field approximation (IMFA), taking into account the short range
order effects (Faleiro Ferreira 1988).
Inspecting table 2, one can see that there are considerable
errors, especially for the strongly anisotropic systems.

The MFA can be considerably improved by placing the clusters in
the mean field instead of separate spins.
A linear chain approximation (LCA) considered by Stout and
Chisholm (1962) when the one-dimensional Ising system is taken
as a cluster leads to the equation
$$
\indeqn{\left(\frac{kT_{\rm c}}{J_{\rm z}}\right)_{\rm LCA}
=2\eta\exp\left(\frac{2J_{\rm z}}{kT_{\rm c}}\right)\eq(16)}
$$
where $\eta=(J_{\rm x}+J_{\rm y})/J_{\rm z}$.
De Bruijn (1958) has shown the solution of an equation like
(16) is given by
$$
\indeqn{\left(\frac{kT_{\rm c}}{J_{\rm z}}\right)_{\rm asympt}
=2/[\ln\eta^{-1}-\ln\ln\eta^{-1}
+O\left(\frac{\ln\ln\eta^{-1}}{\ln\eta^{-1}}\right)]\eq(17)}
$$
when $\eta\to0$.
Fisher (1967) established that the formula (17) is asymptotically
exact for the Ising model.
Although it qualitatively describes the logarithmically slow
drop of the critical temperature with an increase of coupling
anisotropy, unfortunately, this asymptotical formula does not
provide the acceptable precision, even at high anisotropies.
For example, by $J_{\rm x}=J_{\rm y}=10^{-2}J_{\rm z}$ a error of
deviation from the high-temperature series estimate equals
$21\%$ and a error is $28\%$ for the two-dimensional model
($J_{\rm y}=0$) for the same value of anisotropy
($J_{\rm x}/J_{\rm z}=10^{-2}$).

During recent years a number of equations have been obtained
for the critical temperature of a fully anisotropic
three-dimesional  Ising model within the various generalisations
of mean field theory, as well a variational approach (one from
them --- the IMFA --- we have mentioned already).
Using an extended variational method and taking the sum of linear
Ising chains as an auxiliary Hamiltonian, Faleiro Ferreira and
Silva (1982) have found an equation for $T_{\rm c}$ via the
so-called extended linear chain approximation (ELCA).
The numerical solution of this equation shows that the ELCA
perceptibly improves the LCA (see table 2), however, errors are
still considerable.
For instance, the critical point position for the two-dimensional
isotropic case is overstated by $20\%$.

Another new approach named by Faleiro Ferreira (1989), the
improved linear chain approximation (ILCA), is based
upon the same auxiliary Hamiltonian but with other variational
principle, leads to better results only for the isotropic
three-demensional case (this is observed in table 2).

Da Silva \etal (1984) have given the calculation of a phase
transition temperature for the fully anisotropic three-dimensional
Potts lattice, a particular case of which is an Ising model, by the
real space renormalisation group treatment.
However, due to the finite size of spin blocks used, their estimates
lose accuracy very rapidly with an increase of coupling
anisotropy.
Already only by $J_{\rm x}=J_{\rm y}=10^{-1}J_{\rm z}$,
it follows the overestimate
$kT_{\rm c}/J_{\rm z}=1.4552$ which, even after carrying out the
extrapolation on the rather artificial scheme proposed by authors,
falls to $kT_{\rm c}/J_{\rm z}=1.3986$.
This value surpasses the high-temperature series value by $4.1\%$.

We now appeal to our results.
The application of clusters makes it possible to
take into consideration the specific features of a short range
order and as a result, decrease the calculational error.
Therefore, it is not surprising that the finite size scaling
method with its hierarchy of clusters increasing in
growth, allows us to determine the critical point of Ising
model with more exactness than the approaches discussed above
(see again a table 2).
A uniform convergence (contrary to the ILCA) of the
estimates with the growth of a lattice anisotropy is an
important quality of the approximation.
Let us consider a table 1.
In the two-dimensional isotropic limit ($J_{\rm y}=0$ and
$J_{\rm x}/J_{\rm z}=1$), our calculation fixes the critical
temperature with error $4.3\%$ (in the direction of
overestimation).
For $J_{\rm x}/J_{\rm z}\to0$, this error decreases
continiously.
This can be easily checked by making a comparison with the exact
transition temperature values presented in the next column.
In particular, the value has a $1.8\%$ error by
$J_{\rm x}/J_{\rm z}=10^{-1}$.
A similar situation arises in the three-dimensional case
with $J_{\rm x}=J_{\rm y}$.
Here our estimates are again in excess of true values; the
percentage error drops from the maximum value $3.8\%$ in the
fully isotropic case ($J_{\rm x}=J_{\rm y}=J_{\rm z}$) and, for
comparison, -- to $1.7\%$ for the
$J_{\rm x}/J_{\rm z}=10^{-1}$ case.
A analogous picture seems to be preserved in the intermediate
region $0<J_{\rm y}/J_{\rm x}<1$: by fixing $J_{\rm x}/J_{\rm z}$,
the error smoothly moves between the limited values corresponding
to $J_{\rm y}/J_{\rm x}=0$ and $J_{\rm y}/J_{\rm x}=1$.

\section{Conclusions}
In the present paper the more qualitative estimates of critical
temperature in the fully anisotropic three-dimensional Ising
lattice have been derived.
These estimates yield the upper bound everywhere and by
$J_{\rm y}/J_{\rm x}=const$ the error for
$kT_{\rm c}/J_{\rm z}$ monotonically
tends to zero when $J_{\rm x}/J_{\rm z}\to0$.

The accurate analytical solution has been obtained
for an Ising model on the $2\times2\times\infty$ lattice with
fully anisotropic couplings.

The quasi-diagonalisation has been carried out for the transfer
matrix of $3\times3\times\infty$ Ising model with the
rectangular cross section.
The expressions for the matrix elements of subblocks containing
the leading eigenvalues are given in detail.
This permits one to easily reproduce the results presented in the
article, as well as can be useful in considering other problems.

\ack
I am grateful to Miss Laurie Giandomenico for her help in
preparing this article.

\vfill\eject
\append{Explicit form of subblocks $V^{(1)}$ and $V^{(2)}$}
The matrices $\bi{V}^{(1)}$ and $\bi{V}^{(2)}$ are symmetrical,
therefore, it is enough to describe their upper triangular parts:
$$
\bi{V}^{(1)}_{ij}=2\sqrt{\frac{n_j}{n_i}}
\left(\sum^5_{s=1}|g^{(i,j)}_s|
\cosh[(2s-1)K_{\rm z}]\right)
\exp[\case{1}{2}(m^a_i+m^a_j)K_{\rm x}
+\case{1}{2}(m^b_i+m^b_j)K_{\rm y}]\eqno(A1)
$$
and
$$
\bi{V}^{(2)}_{ij}=2\sqrt{\frac{n_j}{n_i}}
\left(\sum^5_{s=1}g^{(i,j)}_s
\sinh[(2s-1)K_{\rm z}]\right)
\exp[\case{1}{2}(m^a_i+m^a_j)K_{\rm x}
+\case{1}{2}(m^b_i+m^b_j)K_{\rm y}]\eqno(A2)
$$
where $i\le j=1,2,\ldots,18$.
The basis vectors are ordered in a non-decrease of their lengths
$$
\indeqn{n_i=\{2,6,6,12,18,18,18,18,18,36,36,36,36,36,36,36,72,72\}.
\eq(A3)}
$$
The quantities $m^a_i$ and $m^b_i$ have a sense of the reduced
partial energies of spin configurations in $i$th vector.
They are
$$
\indeqn{m^a_i=\{9,9,-3,-3,5,5,1,1,-3,1,1,-3,5,-3,1,-3,1,-3\}\eq(A4)}
$$
and
$$
\indeqn{m^b_i=\{9,-3,9,-3,5,1,5,1,-3,1,-3,1,-3,5,-3,1,1,-3\}.\eq(A5)}
$$
Finally, the weight coefficients $g^{(i,j)}_s$ are given as

\settabs4\columns
\+1,1)\ 0\ 0\ 0\ 0\ 1&
  1,2)\ 0\ 1\ 0\ 0\ 0&
  2,2)\ 0\ -2\ 0\ 0\ 1&
  1,3)\ 0\ 1\ 0\ 0\ 0\cr
\+2,3)\ 3\ 0\ 0\ 0\ 0&
  3,3)\ 0\ -2\ 0\ 0\ 1&
  1,4)\ 0\ 1\ 0\ 0\ 0&
  2,4)\ 3\ 0\ 0\ 0\ 0\cr
\+3,4)\ 3\ 0\ 0\ 0\ 0&
  4,4)\ 3\ -2\ 0\ 0\ 1&
  1,5)\ 0\ 0\ 0\ 1\ 0&
  2,5)\ 2\ 0\ 1\ 0\ 0\cr
\+3,5)\ 2\ 0\ 1\ 0\ 0&
  4,5)\ 4\ 0\ 2\ 0\ 0&
  5,5)\ 0\ 0\ 8\ 0\ 1&
  1,6)\ 0\ 0\ 1\ 0\ 0\cr
\+2,6)\ -2\ 0\ 0\ 1\ 0&
  3,6)\ -1\ 2\ 0\ 0\ 0&
  4,6)\ -2\ 4\ 0\ 0\ 0&
  5,6)\ 0\ 7\ 0\ 2\ 0\cr
\+6,6)\ 6\ 0\ 2\ 0\ 1&
  1,7)\ 0\ 0\ 1\ 0\ 0&
  2,7)\ -1\ 2\ 0\ 0\ 0&
  3,7)\ -2\ 0\ 0\ 1\ 0\cr
\+4,7)\ -2\ 4\ 0\ 0\ 0&
  5,7)\ 0\ 7\ 0\ 2\ 0&
  6,7)\ 5\ 0\ 4\ 0\ 0&
  7,7)\ 6\ 0\ 2\ 0\ 1\cr
\+1,8)\ 1\ 0\ 0\ 0\ 0&
  2,8)\ 0\ 2\ -1\ 0\ 0&
  3,8)\ 0\ 2\ -1\ 0\ 0&
  4,8)\ -4\ 2\ 0\ 0\ 0\cr
\+5,8)\ -5\ 4\ 0\ 0\ 0&
  6,8)\ 4\ -3\ 2\ 0\ 0&
  7,8)\ 4\ -3\ 2\ 0\ 0&
  8,8)\ 4\ -4\ 0\ 0\ 1\cr
\+1,9)\ 1\ 0\ 0\ 0\ 0&
  2,9)\ -2\ 1\ 0\ 0\ 0&
  3,9)\ -2\ 1\ 0\ 0\ 0&
  4,9)\ 0\ 4\ -2\ 0\ 0\cr
\+5,9)\ -5\ 4\ 0\ 0\ 0&
  6,9)\ 6\ -2\ 1\ 0\ 0&
  7,9)\ 6\ -2\ 1\ 0\ 0&
  8,9)\ 8\ 0\ 0\ -1\ 0\cr
\+9,9)\ 4\ -4\ 0\ 0\ 1&
  1,10)\ 0\ 0\ 1\ 0\ 0&
  2,10)\ -1\ 2\ 0\ 0\ 0&
  3,10)\ -1\ 2\ 0\ 0\ 0\cr
\+4,10)\ -3\ 2\ 0\ 1\ 0&
  5,10)\ 0\ 7\ 0\ 2\ 0&
  6,10)\ 5\ 0\ 4\ 0\ 0&
  7,10)\ 5\ 0\ 4\ 0\ 0\cr
\+8,10)\ 6\ -2\ 1\ 0\ 0&
  9,10)\ 4\ -3\ 2\ 0\ 0&
  10,10)\ 11\ 0\ 6\ 0\ 1&
  1,11)\ 0\ 1\ 0\ 0\ 0\cr
\+2,11)\ 1\ -1\ 1\ 0\ 0&
  3,11)\ 3\ 0\ 0\ 0\ 0&
  4,11)\ 2\ -2\ 2\ 0\ 0&
  5,11)\ 6\ 0\ 3\ 0\ 0\cr
\+6,11)\ -4\ 4\ 0\ 1\ 0&
  7,11)\ -3\ 6\ 0\ 0\ 0&
  8,11)\ -4\ 4\ -1\ 0\ 0&
  9,11)\ -5\ 2\ -1\ 1\ 0\cr
\+10,11)\ -8\ 8\ 0\ 2\ 0&
  11,11)\ 9\ -5\ 3\ 0\ 1&
  1,12)\ 0\ 1\ 0\ 0\ 0&
  2,12)\ 3\ 0\ 0\ 0\ 0\cr
\+3,12)\ 1\ -1\ 1\ 0\ 0&
  4,12)\ 2\ -2\ 2\ 0\ 0&
  5,12)\ 6\ 0\ 3\ 0\ 0&
  6,12)\ -3\ 6\ 0\ 0\ 0\cr
\+7,12)\ -4\ 4\ 0\ 1\ 0&
  8,12)\ -4\ 4\ -1\ 0\ 0&
  9,12)\ -5\ 2\ -1\ 1\ 0&
  10,12)\ -8\ 8\ 0\ 2\ 0\cr
\+11,12)\ 10\ -4\ 4\ 0\ 0&
  12,12)\ 9\ -5\ 3\ 0\ 1&
  1,13)\ 1\ 0\ 0\ 0\ 0&
  2,13)\ -1\ 0\ -1\ 1\ 0\cr
\+3,13)\ -2\ 1\ 0\ 0\ 0&
  4,13)\ -4\ 2\ 0\ 0\ 0&
  5,13)\ -5\ 4\ 0\ 0\ 0&
  6,13)\ 2\ -4\ 3\ 0\ 0\cr
\+7,13)\ 6\ -2\ 1\ 0\ 0&
  8,13)\ 4\ -2\ 2\ -1\ 0&
  9,13)\ 5\ -3\ 1\ 0\ 0&
  10,13)\ 12\ -4\ 2\ 0\ 0\cr
\+11,13)\ -7\ 7\ -3\ 1\ 0&
  12,13)\ -12\ 6\ 0\ 0\ 0&
  13,13)\ 3\ -7\ 5\ -2\ 1&
  1,14)\ 1\ 0\ 0\ 0\ 0\cr
\+2,14)\ -2\ 1\ 0\ 0\ 0&
  3,14)\ -1\ 0\ -1\ 1\ 0&
  4,14)\ -4\ 2\ 0\ 0\ 0&
  5,14)\ -5\ 4\ 0\ 0\ 0\cr
\+6,14)\ 6\ -2\ 1\ 0\ 0&
  7,14)\ 2\ -4\ 3\ 0\ 0&
  8,14)\ 4\ -2\ 2\ -1\ 0&
  9,14)\ 5\ -3\ 1\ 0\ 0\cr
\+10,14)\ 12\ -4\ 2\ 0\ 0&
  11,14)\ -12\ 6\ 0\ 0\ 0&
  12,14)\ -7\ 7\ -3\ 1\ 0&
  13,14)\ 10\ -6\ 2\ 0\ 0\cr
\+14,14)\ 3\ -7\ 5\ -2\ 1&
  1,15)\ 1\ 0\ 0\ 0\ 0&
  2,15)\ 0\ 2\ -1\ 0\ 0&
  3,15)\ -2\ 1\ 0\ 0\ 0\cr
\+4,15)\ -2\ 3\ -1\ 0\ 0&
  5,15)\ -5\ 4\ 0\ 0\ 0&
  6,15)\ 4\ -3\ 2\ 0\ 0&
  7,15)\ 6\ -2\ 1\ 0\ 0\cr
\+8,15)\ 3\ -4\ 2\ 0\ 0&
  9,15)\ 3\ -4\ 2\ 0\ 0&
  10,15)\ 10\ -5\ 3\ 0\ 0&
  11,15)\ -10\ 4\ -2\ 2\ 0\cr
\+12,15)\ -8\ 8\ -2\ 0\ 0&
  13,15)\ 8\ -4\ 4\ -2\ 0&
  14,15)\ 10\ -6\ 2\ 0\ 0&
  15,15)\ 7\ -8\ 2\ 0\ 1\cr
\+1,16)\ 1\ 0\ 0\ 0\ 0&
  2,16)\ -2\ 1\ 0\ 0\ 0&
  3,16)\ 0\ 2\ -1\ 0\ 0&
  4,16)\ -2\ 3\ -1\ 0\ 0\cr
\+5,16)\ -5\ 4\ 0\ 0\ 0&
  6,16)\ 6\ -2\ 1\ 0\ 0&
  7,16)\ 4\ -3\ 2\ 0\ 0&
  8,16)\ 3\ -4\ 2\ 0\ 0\cr
\+9,16)\ 3\ -4\ 2\ 0\ 0&
  10,16)\ 10\ -5\ 3\ 0\ 0&
  11,16)\ -8\ 8\ -2\ 0\ 0&
  12,16)\ -10\ 4\ -2\ 2\ 0\cr
\+13,16)\ 10\ -6\ 2\ 0\ 0&
  14,16)\ 8\ -4\ 4\ -2\ 0&
  15,16)\ 11\ -4\ 2\ -1\ 0&
  16,16)\ 7\ -8\ 2\ 0\ 1\cr
\+1,17)\ 0\ 1\ 0\ 0\ 0&
  2,17)\ 1\ -1\ 1\ 0\ 0&
  3,17)\ 1\ -1\ 1\ 0\ 0&
  4,17)\ 4\ -1\ 1\ 0\ 0\cr
\+5,17)\ 6\ 0\ 3\ 0\ 0&
  6,17)\ -4\ 4\ 0\ 1\ 0&
  7,17)\ -4\ 4\ 0\ 1\ 0&
  8,17)\ -5\ 2\ -1\ 1\ 0\cr
\+9,17)\ -4\ 4\ -1\ 0\ 0&
  10,17)\ -7\ 10\ 0\ 1\ 0&
  11,17)\ 10\ -4\ 4\ 0\ 0&
  12,17)\ 10\ -4\ 4\ 0\ 0\cr
\+13,17)\ -7\ 7\ -3\ 1\ 0&
  14,17)\ -7\ 7\ -3\ 1\ 0&
  15,17)\ -9\ 6\ -2\ 1\ 0&
  16,17)\ -9\ 6\ -2\ 1\ 0\cr
\+17,17)\ 19\ -9\ 7\ 0\ 1&
  1,18)\ 1\ 0\ 0\ 0\ 0&
  2,18)\ -2\ 1\ 0\ 0\ 0&
  3,18)\ -2\ 1\ 0\ 0\ 0\cr
\+4,18)\ -3\ 1\ -1\ 1\ 0&
  5,18)\ -5\ 4\ 0\ 0\ 0&
  6,18)\ 6\ -2\ 1\ 0\ 0&
  7,18)\ 6\ -2\ 1\ 0\ 0\cr
\+8,18)\ 5\ -3\ 1\ 0\ 0&
  9,18)\ 4\ -2\ 2\ -1\ 0&
  10,18)\ 8\ -6\ 4\ 0\ 0&
  11,18)\ -7\ 7\ -3\ 1\ 0\cr
\+12,18)\ -7\ 7\ -3\ 1\ 0&
  13,18)\ 10\ -6\ 2\ 0\ 0&
  14,18)\ 10\ -6\ 2\ 0\ 0&
  15,18)\ 9\ -5\ 3\ -1\ 0\cr
\+16,18)\ 9\ -5\ 3\ -1\ 0&
  17,18)\ -19\ 13\ -3\ 1\ 0&
  18,18)\ 13\ -13\ 7\ -2\ 1&&\cr

Using (A1)--(A5) and above values for the weight coefficients,
one can easily find the expressions for the matrix elements.
For instance,
$$
\indeqn{\bi{V}^{(1)}_{11}=2\cosh(9K_{\rm z})\exp[9(K_{\rm x}
+K_{\rm y})]}
$$
$$
\indeqn{\bi{V}^{(1)}_{35}=2\sqrt3[2\cosh(K_{\rm z})
+\cosh(5K_{\rm z})]\exp(K_{\rm x}+7K_{\rm y})}
$$
and
$$
\eqalignno{\bi{V}^{(2)}_{18,18}&=2\{13[\sinh(K_{\rm z})
-\sinh(3K_{\rm z})]+7\sinh(5K_{\rm z})-2\sinh(7K_{\rm z})\cr
&\phantom{=}{}+\sinh(9K_{\rm z})\}\exp[-3(K_{\rm x}
+K_{\rm y})].\cr}
$$

\references
\refbk{Barber M N 1983}{Phase Transitions and Critical Phenomena}{vol 8,
ed C Domb and J L Lebowitz (New York: Academic) p 145}
\refbk{de Bruijn N G 1958}{Asymptotic Methods in Analysis}{(Amsterdam:
North--Holland) sections 2.4 and 2.7}
\refjl{Faleiro Ferreira J R and Silva N P 1982}{\PSS\ \rm b}{114}{47}
\refjl{Faleiro Ferreira J R 1988}{\PSS\ \rm b}{150}{281}
\refjl{\dash 1989}{\PSS\ \rm b}{156}{647}
\refjl{Ferrenberg A M and Landau D P 1991}{\PR\ \rm B}{44}{5081}
\refjl{Fisher M E 1967}{\PR}{162}{480}
\refjl{Kaufman B 1949}{\PR}{76}{1232}
\refjl{Livet F 1991}{Europhys. Lett.}{16}{139}
\refjl{Navarro R and de Jongh L J 1978}{Physica\ \rm B}{94}{67}
\refjl{Nightingale M P 1976}{Physica\ \rm A}{83}{561}
\refjl{\dash 1982}{\JAP}{53}{7927}
\refjl{Onsager L 1944}{\PR}{65}{117}
\refjl{da Silva L R, Tsallis C and Schwachheim G 1984}{\JPA}{17}{3209}
\refbk{Sominskii I S 1967}{Elementarnaya Algebra. Dopolnitelny
Kurs}{(Moskva: Nauka) chap 1}
\refjl{Stout J W and Chisholm R C 1962}{\JCP}{36}{979}
\refjl{Yurishchev M A 1989}{\PSS\ \rm b}{153}{703}
\refjl{Yurishchev M and Sterlin A 1991}{J. Phys.: Condens. Matter}
{3}{2373}

\tables

\tabcaption{Normalized critical temperature $kT_{\rm c}/J_{\rm z}$
for the fully anisotropic three-dimensional Ising model as a
function of $J_{\rm x}/J_{\rm z}$ and $J_{\rm y}/J_{\rm x}$.}

\medskip
\boldrule{104mm}
\halign{&#\hfil\cr
&\hfil $J_{\rm y}/J_{\rm x}$\span\omit\span\omit
\span\omit\span\omit\span\omit\cr
\noalign{\vskip-6pt}
$J_{\rm x}/J_{\rm z}$\qquad&\hrulefill\span\omit\span\omit
\span\omit\span\omit\span\omit\cr
&0.0\qquad&\qquad&0.25\qquad&0.50\qquad&0.75\qquad&1.0\cr
\noalign{\medrule{104mm}}
1.0&2.367$\;\;$2.269\dag&&3.277&3.819&4.275&4.685$\;\;$4.5106\ddag\cr
0.9&2.246$\;\;$2.153\dag&&3.079&3.575&3.993&4.368\cr
0.8&2.120$\;\;$2.034\dag&&2.876&3.325&3.704&4.045\cr
0.7&1.987$\;\;$1.909\dag&&2.667&3.070&3.409&3.714\cr
0.6&1.848$\;\;$1.779\dag&&2.451&2.806&3.105&3.375\cr
0.5&1.699$\;\;$1.641\dag&&2.226&2.533&2.791&3.024$\;\;$2.9286\ddag\cr
0.4&1.540$\;\;$1.492\dag&&1.988&2.247&2.464&2.659$\;\;$2.580\ddag\cr
0.3&1.365$\;\;$1.328\dag&&1.733&1.943&2.117&2.273$\;\;$2.219\ddag\cr
0.2&1.167$\;\;$1.141\dag&&1.450&1.608&1.738&1.854$\;\;$1.814\ddag\cr
0.1&0.921$\;\;$0.905\dag&&1.109&1.211&1.293&1.366$\;\;$1.343\ddag\cr
0.09&0.891$\;\;$0.877\dag&&1.069&1.164&1.241&1.309\cr
0.08&0.859$\;\;$0.846\dag&&1.026&1.115&1.187&1.251\cr
0.07&0.826$\;\;$0.814\dag&&0.982&1.064&1.131&1.189\cr
0.06&0.790$\;\;$0.779\dag&&0.934&1.010&1.070&1.124\cr
0.05&0.751$\;\;$0.741\dag&&0.882&0.951&1.006&1.054$\;\;$1.041\ddag\cr
0.04&0.707$\;\;$0.698\dag&&0.825&0.886&0.935&0.978\cr
0.03&0.657$\;\;$0.650\dag&&0.760&0.813&0.856&0.892\cr
0.02&0.597$\;\;$0.590\dag&&0.683&0.727&0.761&0.791\cr
0.01&0.513$\;\;$0.508\dag&&0.579&0.611&0.637&0.658$\;\;$0.65\ddag\cr
0.009&0.502$\;\;$0.498\dag&&0.565&0.596&0.621&0.641$\;\;$0.637\S\cr
0.008&0.491$\;\;$0.486\dag&&0.551&0.581&0.604&0.624$\;\;$0.619\S\cr
0.007&0.478$\;\;$0.474\dag&&0.535&0.564&0.586&0.604$\;\;$0.600\S\cr
0.006&0.464$\;\;$0.460\dag&&0.518&0.545&0.566&0.583$\;\;$0.579\S\cr
0.005&0.449$\;\;$0.445\dag&&0.499&0.524&0.544&0.560$\;\;$0.556\S\cr
0.004&0.431$\;\;$0.428\dag&&0.478&0.501&0.519&0.534$\;\;$0.531\S\cr
0.003&0.410$\;\;$0.407\dag&&0.453&0.474&0.490&0.503$\;\;$0.500\S\cr
0.002&0.383$\;\;$0.380\dag&&0.421&0.439&0.453&0.465$\;\;$0.462\S\cr}
\boldrule{104mm}

\noindent\dag\ Onsager (1944).

\noindent\ddag\ Navarro and de Jongh (1978).

\noindent\S\ Yurishchev and Sterlin (1991).

\vfill\eject
\tabcaption{Critical temperature estimates of the fully anisotropic
three-dimensional Ising model versus calculational method.}

\medskip
\boldrule{113mm}
\halign{&#\hfil\qquad\cr
&\hfil$J_{\rm x}/J_{\rm z}=1$\hfil\span\omit\qquad&\hfil
$J_{\rm x}/J_{\rm z}=10^{-2}$\hfil\span\omit\cr
\noalign{\vskip-6pt}Method\qquad&\hrulefill\span\omit
\qquad&\hrulefill\span\omit\cr
&$J_{\rm y}=0$\qquad&$J_{\rm y}=J_{\rm x}$
\qquad&$J_{\rm y}=0$\qquad&$J_{\rm y}=J_{\rm x}$\cr
\noalign{\medrule{113mm}}
MFA&4&6&2.02&2.04\cr
IMFA&3.230&4.933&1.465&1.487\cr
LCA&3.526&5.686&0.590&0.699\cr
ILCA&2.885&4.622&0.588&0.695\cr
ELCA&2.728&4.881&0.543&0.669\cr
Table 1&2.367&4.685&0.513&0.658\cr
Exact&2.2691&4.5115&0.5089&0.65\cr}
\boldrule{113mm}

\bye